\documentclass{aa}

\newcommand\MSun{M_{\odot} }

\newcommand\Rg{R_{\rm g}}
\newcommand\Rin{R_{\rm in}}

\newcommand\grs{GRS 1915+105 }
\newcommand\dbb{\textit{diskbb} }

\newcommand\thcds{\textit{thComp} }

\newcommand\bb{\textit{bbody} }
\newcommand\cps{\textit{compPS} }

\newcommand\Ka{K$_{\alpha}$\ }

\usepackage{epsfig, amssymb}

\begin{document}

\title{Spectral modeling of the three spectral states of the Galactic microquasar GRS 1915+105}

\authorrunning{M. A. Sobolewska \& P. T. \.{Z}ycki}

\titlerunning{X--ray modeling of GRS 1915+105}

\offprints{M. A. Sobolewska} 

\author{Ma\l{}gorzata A. Sobolewska \and Piotr T. \.{Z}ycki}

\institute{Nicolaus Copernicus Astronomical Center, Bartycka 18, 00-716 Warsaw, Poland \\
           \email{(malsob, ptz)@camk.edu.pl}}

 \date{Received ...; accepted ...}

\abstract {

We have analyzed Rossi X--ray Timing Explorer spectral data of the Galactic micro-quasar
GRS 1915+105 in its various spectral states, as defined by Belloni et al.\ (2000).
In states A and B the spectra
are dominated by a strong soft thermal component, accompanied by a weak harder tail.
The soft component is rather complex and  cannot be described as a simple accretion
disk emission. Relativistic effects in Kerr metric contribute to the
complexity of the soft component but are not sufficient to fully account for it.
As found previously, state C spectra are dominated by a Comptonized component, 
with small contribution from disk photons. 
The X--ray reprocessed component is highly
significant in those spectra and, in contrast to the usual hard state spectra
from accreting black holes, it is highly ionized. 

}

\maketitle

\section{Introduction}

Ever since its discovery by Castro-Tirado et al.\ (\cite{cas94}), the transient Galactic source 
GRS 1915+105 has been a subject of intense observational and theoretical work.
The source shows a number of remarkable features, most spectacular of which are ejections 
of relativistic jets with apparent super-luminal motion (Mirabel \& Rodriguez \cite{mir94}).
It displays a large variety of behaviour
in the X-ray band (dramatic flux and spectral variability on time scales from
miliseconds to months), most likely reflecting violent phenomena occurring in the inner
accretion flow.

Large amount of data collected by various X-ray observatories indicate that the source
stays permanently in a soft spectral state (see, e.g., Nowak \cite{now95} for 
a characteristics of
the various spectral/timing states). That is, its spectral energy distribution
is dominated by the 1--10 keV band, even though the ratio of soft thermal 
(temperature $\sim 1$ keV) emission to harder Comptonized continuum does vary
substantially. The source never seems to have been in the hard spectral state, which
is more typical for black hole binaries with energy distribution peaking at $\sim 100$ keV
(see Poutanen \cite{pou98} for review).

A characteristic feature of \grs is its strong variability, of both flux (count rate)
and energy spectra. Among the variety of light curves and color--color diagrams certain
patterns can be found. These were studied in detail by Belloni et al.\ (2000; hereafter 
\cite{bel00}),  using {\it RXTE\/} data, who
attempted to  classify the behaviour of the source. They divided the lightcurves 
into 12 classes based on the color-color diagrams and count rates, and found 
that the lightcurves from each class can be decomposed into only 3 basic spectral states
(A, B, C in the nomenclature of \cite{bel00}). 
The strongly variable lightcurves of the source could be explained in terms of rapid switches 
between these
three states. 

There have been previous attempts to model the X--ray spectra of \grs in the B and C spectral
states. Vilhu et al.\ (\cite{vil01}) analyzed {\it RXTE\/} data (PCA/HEXTE instruments, 
in the range of 2-60 keV) and concluded
that spectra of GRS 1915+105 can be described with the model consisting of a disc blackbody and a
Comptonized component with Compton reflection. Zdziarski et al.\ (\cite{zdz01}) presented an
analysis of simultaneous observations of \grs in states B and C  by {\it RXTE\/} and 
{\it CGRO\/}/OSSE instruments. 
They found that the broad-band spectra  required non-thermal
Comptonization of soft photons, since the data do not show any break up to at least 600 keV.
Low temperature, thermal Comptonization was also needed  to correctly model (together with blackbody component) the soft X--ray
part ($E<10$ keV), thus the best final model invoked a {\it hybrid\/} plasma 
(Coppi \cite{cop99}).
Vadavale et al.\ (\cite{vad01})
focused on study of the source during the steady low-hard state (class
$\chi$, state C), which they further subdivided into radio-loud ($\chi_{RL}$) and
radio-quiet ($\chi_{RQ}$) ones. From their analysis followed that the spectrum of
\grs in state C can be described as a superposition of disk blackbody and
Comptonized component, and an additional power low component in the case of the
$\chi_{RL}$ state. This additional power low was interpreted as an emission from the
base of a jet.
Analysis of spectral features due to iron emission/absorption in the 5--9 keV band
was performed by Martocchia et al.\ (\cite{mar02}) using BeppoSAX data of the source in state C,
according to their interpretation. They found  that
the Fe \Ka line is broad and it has the characteristic asymmetric profile indicating
its origin from a rotating accretion disk.

In all above mentioned analyses, disk blackbody or blackbody component is used to (fully or partially)
describe the soft emission from the disk. The \textit{diskbb\/}
model reproduces the shape of the disk spectrum quite accurately unless the relativistic 
corrections are extreme. However, the numerical values of the parameters
do not easily translate into physical parameters (Merloni et al.\ \cite{mer00}). 
These can be estimated from
a relativistic model, which includes  radiative transfer effects, at
least approximating them by introducing  the color temperature correction
(Shimura \& Takahara \cite{shi95}). 

Therefore, in this paper we analyse data from all three spectral states of \grs,
applying the same set of models to all the data. The models include one
for emission from an accretion disk around a maximally rotating Kerr black hole.

\section{Data selection and reduction}

The log of PCA/HEXTE observations used for our analysis is given in Table \ref{log}.
The data were extracted from public archive at HEASARC/GSFC and reduced with the FTOOLS
software package (version 5.0).

Based on \cite{bel00} we attempted to extract spectra corresponding to 'pure' spectral states
A, B and C. We have two observations from class $\chi$ (pure state C),
one observation from class $\lambda$ from which we extract a period corresponding to 
state B based on the X--ray color--color
diagrams and lightcurves appearance (see \cite{bel00} for details), one observation from class
$\theta$ which provides state A spectra, and two observations from class $\beta$ --
further examples of state A, B, and C spectra.

For the spectral analysis, PCA \textit{Standard 2} mode data (PCU 0-2, top
layer only) and HEXTE \textit{Archive Mode} (both clusters) configuration are used.
The energy range is 3-20 keV and 20-100 keV, correspondingly. We assume systematic
errors in the PCA data at the level of 1\% due to the uncertainty of response
matrices. Standard background subtraction and dead-time correction procedures are
applied.

The value of the hydrogen column density, $N_H$, is allowed to vary during the fit. We
assume the source inclination is  $i=70^\circ$, distance $d=12.5$ kpc 
(Mirabel \& Rodriguez \cite{mir94}) 
and central black hole mass $M=14 \MSun$ (Greiner et al.\ \cite{gre01}).
We allow for free relative normalization of 
HEXTE data with respect to PCA data. Analysis was performed in {\sc XSPEC} ver.\ 11
(Arnaud \cite{arn96}).


\begin{table}
\begin{center}
\caption{\label{log}
The log of observations. The letters I, K stand for observations ID 10408-01, 20402-01,
respectively (\cite{bel00}). The third column indicates which state spectra were extracted from
the data.}
\begin{tabular}{ c c c }\hline\hline
Observation ID & Class & State \\ \hline 

K-44-00	& $\beta$	& A, B, C \\
K-45-03	& $\beta$	& A, B, C \\
K-45-02 & $\theta$      & A, B    \\
I-38-00	& $\lambda$	& B       \\
K-05-00	& $\chi^{2}$	& C	  \\
I-42-00	& $\chi^{4}$	& C	  \\ \hline

\end{tabular}
\end{center}
\end{table}


\section{Models}

The disk soft thermal emission will be described by the simple disk blackbody model 
(Mitsuda et al.\ \cite{mit84};  implemented as
{\it diskbb\/}) in XSPEC, or a relativistic accretion disk model for Kerr geometry 
(Ebisawa et al.\ \cite{ebi01}). 
The {\it KerrD\/} model is computed for the maximally rotating black hole
($a=0.998$; Thorne \cite{thorne74}). It assumes blackbody spectra at each radius with the color
temperature correction constant with radius and convolves them with
the transfer function describing photon propagation, as computed by Laor (\cite{laor91}).
We fix the black hole mass, source distance and accretion disk
inclination 
at the values indicated above, and allow only
the inner  radius of the disk (in units of $\Rg \equiv GM/c^2$) and the mass accretion rate,
$\dot M$ to be free parameters. In particular, the model does not have an additional 
free normalization.

The Comptonized component will be described by the {\it thComp\/} model, based on a 
solution of 
the Kompaneets equation (Zdziarski et al.\ \cite{zjm96}), or 
by the {\it compPS\/} model of Poutanen \& Svensson (\cite{pou96}).
The latter does not assume  the diffusion approximation but solves for each scattering order
independently. The {\it compPS\/} model can be used with purely thermal, purely non-thermal 
or a  hybrid electron energy distribution (Coppi \cite{cop99}). 
The seed photons will be assumed to have a disk blackbody spectrum and the comptonizing cloud
is assumed to be a sphere.
The code also consistently computes the un-scattered fraction of seed photons. 

\begin{table*}

\caption[]{Values of $\frac{\chi^2}{d.o.f.}$ ($\chi_{\nu}^2$) for various models fitted
to the data.}
\label{tab:chi2}
\begin{center}
\renewcommand*{\arraystretch}{1.3}

\begin{tabular}{ l l  l  l  l  l } 
\hline\hline

State & Obs. Id &            (1)~~~~~~~~~~ & (2)~~~~~~~~~~ &    (3)~~~~~~~~~~ & (4)~~~~~~~~~~\\ \hline

   &  K-44-00 & $\frac{114.5}{101}$ (1.14) &  --           &    --            &         --    \\
C  & K-45-03     & $\frac{84.2}{101}$ (0.83)  &  --        &    --            &         --     \\
   & K-05-00     & $\frac{83.8}{96}$ (0.87) &    --        &    --            &         --     \\
   & I-42-00     & $\frac{102.9}{99}$ (1.04) &   --        &    --            &         --     \\

\hline

   &  K-44-00 & 4.8 & $\frac{45.6}{65}$ (0.70) & $\frac{44.6}{66}$ (0.68) & $\frac{43.8}{66}$ (0.66) \\
A  & K-45-03     & 3.4 & $\frac{75.1}{61}$ (1.23) & $\frac{73.7}{62}$ (1.19) & $\frac{79.9}{62}$ (1.29) \\
   & K-45-02     & 3.4 & $\frac{54.1}{65}$ (0.83) & $\frac{54.8}{66}$ (0.83) & $\frac{63.3}{66}$ (0.96) \\

\hline

   &  K-44-00 & 2.7 & $\frac{51.8}{66}$ (0.79) & $\frac{47.1}{67}$ (0.70) & $\frac{43.3}{67}$ (0.65) \\
B  & K-45-03     & 1.6 & $\frac{85.4}{85}$ (1.00) & $\frac{83.5}{86}$ (0.97) & $\frac{80.0}{86}$ (0.93) \\
   & I-38-00     & 2.3 & $\frac{88.2}{95}$ (0.93) & $\frac{85.4}{96}$ (0.89) & $\frac{81.0}{96}$ (0.84) \\

\hline

\end{tabular}
\end{center}

Models (2), (3) and (4) fitted to state A spectra contain an additional gaussian line

Fitted models: \\
(1) \cps -- comptonized disk blackbody \\
(2) \thcds+\cps -- two comptonized disk blackbody spectra\\
(3) \bb+\cps -- comptonized disk blackbody with an additional blackbody\\
(4) {\it KerrD\/}+\cps -- Kerr metric disk spectrum and a comptonized disk blackbody\\
In all fits the \cps model includes the reprocessed component
\end{table*}


For the X-ray reprocessed component
we will use the {\it relrepr\/} model (\.{Z}ycki et al.\ \cite{zyc99}), combining the
Compton-reflected continuum (Magdziarz \& Zdziarski \cite{mag95}; Done et al.\ \cite{done92}) 
with the Fe \Ka line 
computations of \.{Z}ycki \& Czerny (\cite{zyc94}). The model computes self-consistently the
strength of the line for a given primary continuum spectrum, 
amplitude of the Compton-reflected component, $R$, ionization
parameter, $\xi=4 \pi F_X/n$, and Fe abundance, and it also allows for relativistic effects
to be applied to both components (using prescription from Fabian et al.\ \cite{frsw89}).
The amplitude $R$ is defined as $\Omega/(2\pi)$, where $\Omega$ is the solid angle
subtended by the reprocessor from the X--ray source.

\section{Results}

\subsection{State B spectra}

We begin  our analysis with the brightest state, B. First, 
we try modeling soft part of the spectrum with disk blackbody, and the hard one
with non-thermal Comptonization (we apply {\it compPS\/} model to the data).
The latter is motivated by the reports, where
the source spectrum in state B and C was found to extend to $\ge 600$ keV without a break 
or a cutoff (Zdziarski et al.\ \cite{zdz01}).  We assume that the seed photons for Comptonization
have the same disk blackbody shape and temperature.
The reprocessed component is also included in the model.
This model gives unacceptable fits, $\chi^2_\nu \sim 2$, with residuals up
to 10 per cent (Model 1 in Table~\ref{tab:chi2}).
Good description of data is provided by models including a more complex shape of
the soft component. Motivated by earlier results (see \.{Z}ycki et al.\ \cite{zyc01};
 Wilson \& Done \cite{wil01} and references therein) 
we test two phenomenological descriptions of the soft 
component: (i) with an additional Comptonized disk blackbody 
 and (ii) an additional  blackbody. 
 (Note that a certain fraction of the seed disk blackbody photons always contributes 
to the soft component
due to modeling the hard part of the spectrum with {\it compPS\/}.) 
Both these complex models give significant improvement compared to Model 1,
with best fits of comparable quality, as shown in 
Table~\ref{tab:chi2} (Models 2 and 3).

A possible physical reason for the complexity may be the relativistic effects expected
from an inclined accretion disk. Therefore we also try (iii) the {\it KerrD\/}
model for the soft component (with non-thermal {\it compPS\/} for 
the hard tail, as before). The quality of the fits is similar for all the three
models (Model 4 in Table~\ref{tab:chi2}).
The relativistic effects are quite pronounced at the high inclination of the source, 
$i\approx 70^\circ$ (see Ebisawa et al.\ \cite{ebi01}; Gierli\'{n}ski et al.\ \cite{gie01}), 
and the Doppler-boosted disk emission fits well the observed high energy cutoff 
of the soft component.
In the best fit for K-44-00 te inner disk radius $\Rin = 3.80^{+2.20}_{-2.56}\,\Rg$,
for $i=70^{\circ}$ and $d=12.5$ kpc. With the more recently 
determined $i=66^{\circ}$ and $d=11$ kpc (Fender et al.\ \cite{fender99}) the inner radius
is $\Rin = 2.25^{+1.0}_{-1.01}$. In both cases $\Rin$ is consistent with the last stable
orbit at $1.24\,\Rg$, and a large inner hole in the disk is ruled out.
A good fit ($\chi^2/\nu = 46.3/72$) can be obtained with the disk emission model
in Schwarzschild metric, however a small mass of the black hole, $M\approx 4 \MSun$
is required in this case, in order to account for the high observed disk temperature.
Thus, if the soft component is modeled as a relativistic disk emission, a Kerr (rotating) black 
hole is implied by our result. 

\begin{figure*}
 \parbox{1.0\textwidth}{
  \hfil \parbox{0.32\textwidth}{
   \epsfxsize=0.3\textwidth \epsfbox{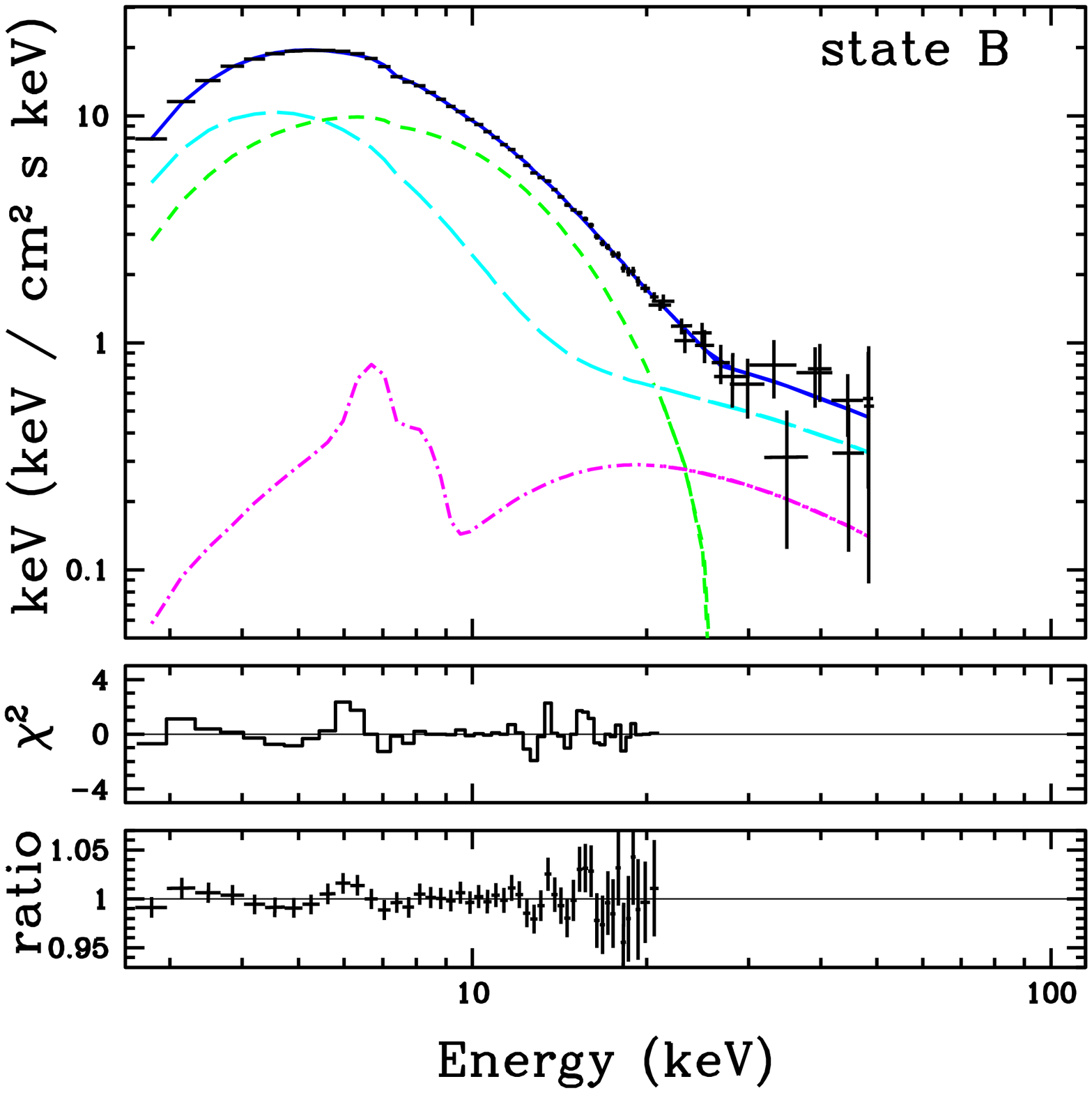}}
  \hfil 
  \parbox{0.32\textwidth}{
   \epsfxsize=0.3\textwidth \epsfbox{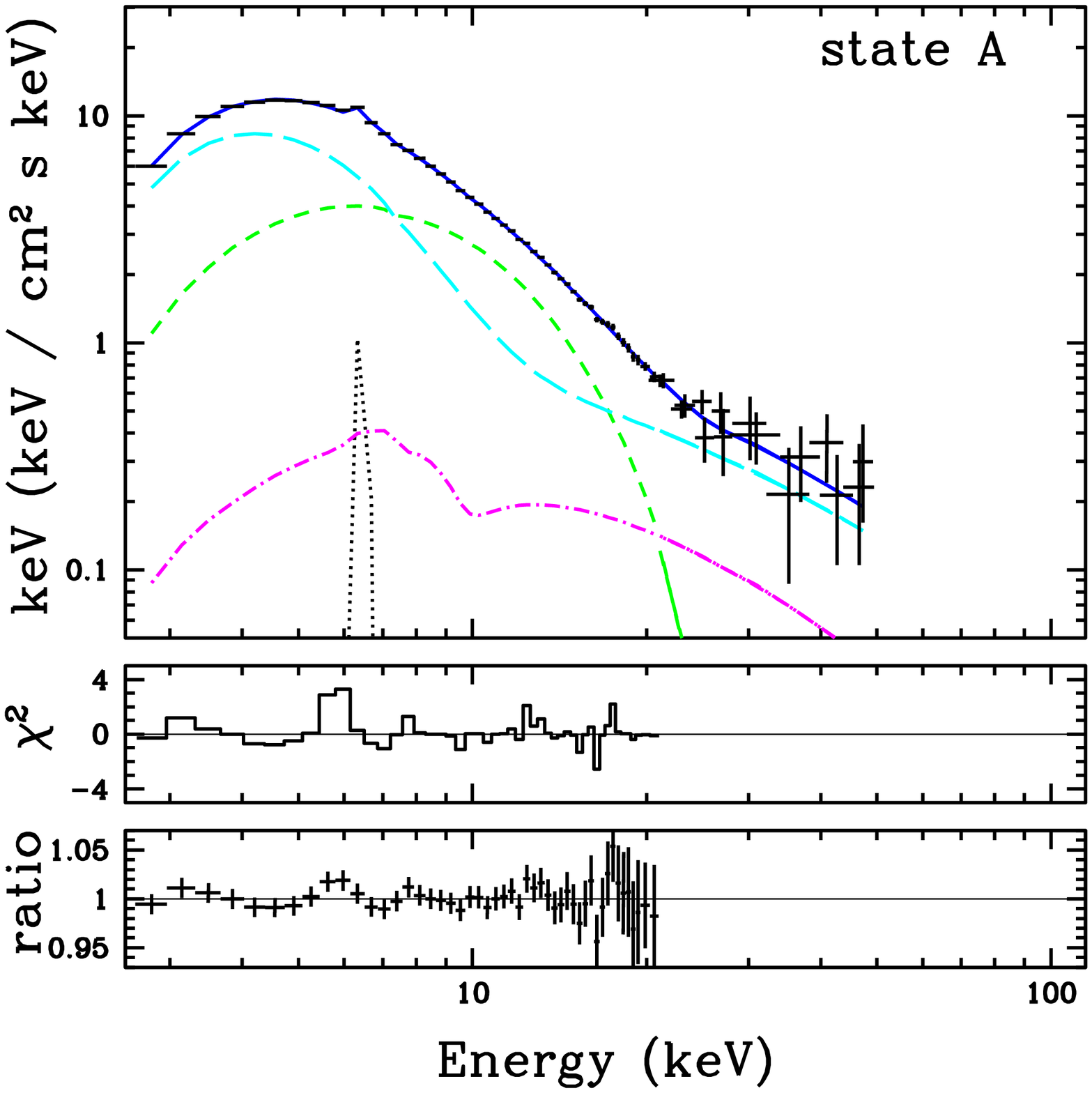}}
  \hfil
  \parbox{0.32\textwidth}{
   \epsfxsize=0.3\textwidth \epsfbox{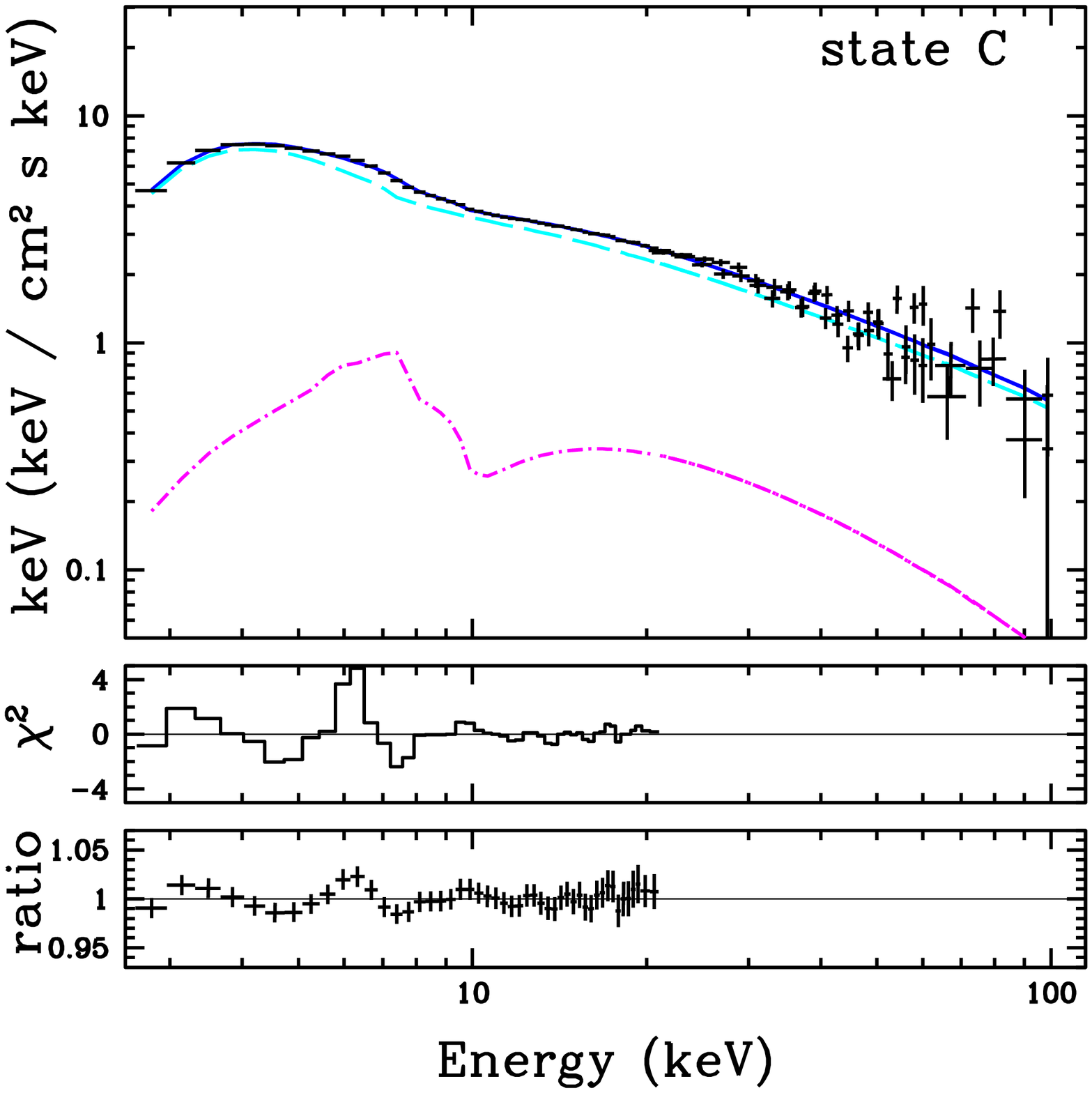}}
}
\caption{{\it KerrD\/} model fits to spectra from observation K-44-00. Long-dashed curves are the 
comptonized component {\it compPS\/} with the transmitted seed photons, Short-dashed curves
are the Kerr disk spectrum model, and the dash-dot curves are the reprocessed component.
Complexity of the soft component in states A and B is apparent and this cannot be accounted
for by the Kerr disk model alone. There is only a weak disk thermal component in state C
spectrum, but modeling it with the {\it KerrD\/} model suggests that the disk $\dot M$ may
actually be comparable or even higher than in states A and B. However, the inner disk radius
is larger than the last stable orbit, so most of the accretion power would not appear
as thermal X--ray emission.
\label{fig:44specs}
}
\end{figure*}

We note once again that some seed disk blackbody photons do contribute to the soft
component (Fig.~\ref{fig:44specs})
since the {\it compPS\/} model computes self-consistently the un-scattered
fraction of the seed photon flux. Without this component the fit is worse by 
$\Delta\chi^2 = 21$ (for K-44-00), therefore we stress that the
relativistic effects may be not the only factor contributing to the complexity
of the soft component. However, given the high inclination of the accretion disc,
their {\it a priori\/} importance is obvious.
The mass accretion rate is $7.5\times10^{18}\,{\rm g\,s^{-1}}$ for K-44-00 data
(Table~\ref{tab1}; and up to $2.6\times10^{19} \,{\rm g\,s^{-1}}$ for other data).
We note that both $\dot M$ and $\Rin$ would be different
for different value of the black hole angular momentum (see discussion in 
Gierli\'{n}ski et al.\ \cite{gie01};
recall that $a=0.998$ is assumed in our paper), 
thus our result can only be regarded as a qualitative demonstration that relativistic 
effects can partially account for apparently complex soft component in this spectral state.

The seed photon temperature for the non-thermal Comptonization was a free parameter
in these fits, but the best fit values are similar to the obtained maximum temperatures 
of the Kerr disk emission ($\approx 1.2$ keV), as expected. 
The reprocessed component from the non-thermal
primary is present in the data, with typical amplitude $R=0.7$--$1.0$ (Table~\ref{tab1}). 
Its presence is however not highly significant, since removing it we obtain fits worse by
$\Delta\chi^2 = 5.4$--$8.4$.  The best fit value of ionization parameter is high,
$\xi \sim 10^4$, as expected from a hot accretion disk, although the formal errors
on $\xi$ are large. No further broadening of the reprocessed component is formally
required by the data, although the data are consistent with relativistic smearing
from a disk extending down to $\sim 20\,\Rg$.
This is larger than 
the inner radius of the emitting disk, indicating that the innermost disk region
is too hot to produce  any significant spectral features. A part of the uncertainty
in determining $\Rin$ comes from the fact that 
Compton scattering in highly ionized plasma contributes to smearing of the spectral
features, making quantitative determinations of $\Rin$ rather uncertain 
(Ross et al.\ \cite{rfy99}).

\begin{table*}
\caption{\label{tab1}
Fit parameters and $\chi^2$ values for observation K-44-00 in states A, B, and C. 
Model consists of \textit{KerrD\/} and non-thermal \textit{compPS} (A and B), or single \textit{compPS\/} (C)
}
\begin{center}
\renewcommand*{\arraystretch}{1.2}
\begin{tabular}{ c  c c c } 
\hline
Parameter & State B & State A & State C \\ \hline
\bf $n_{H}$ [$\times 10^{22}$ cm$^{-2}$]  & \bf 6.2$^{+0.3}_{-0.2}$	  & \bf 5.9$^{+0.2}_{-0.3}$	    & \bf 5.6$^{+0.7}_{-0.5}$  \\ \hline
~~\.M  [$\times 10^{18}$ gs$^{-1}$]~~ & 7.5$^{+4.5}_{-3.0}$       & 1.43$^{+1.50}_{-0.02}$                  & - \\
~$R_{\rm in}$  [$R_{\rm g}$]          & 3.80$^{+2.20}_{-2.56}$    & 1.47$^{+2.50}_{-0.23}$                  & -\\ \hline
~$kT_{\rm in}$  (keV)                 & 1.30$^{+0.13}_{-0.26}$    & 1.16$^{+0.03}_{-0.03}$                  & 0.94$^{+0.06}_{-0.08}$ \\
~$\Gamma_e$                           & 4.8$^{+5.1}_{-0.9}$       & 5.9$^{+1.1}_{-1.7}$                     & 6.7$^{+0.8}_{-0.5}$\\
~$\tau$                               & 0.05$^{+0.10}_{-0.05}$    & 0.06$^{+0.04}_{-0.04}$                  & 0.44$^{+0.11}_{-0.22}$ \\
~R=$\frac{\Omega}{2\pi}$              & 1 (fixed)                 & 0.70$^{+\infty}_{-0.58}$                & 0.30$^{+0.20}_{-0.15}$ \\
~$\xi$                                &~~2200$^{+17000}_{-2000}$~~&~~1.10$^{+\infty}_{-1.02}$$\times$10$^4$~~&~~6200$^{+23000}_{-6000}$~~\\
~$R_{\rm in, refl}$ [$R_{\rm g}$]     & 165$^{+\infty}_{-145}$    & 137$^{+\infty}_{-131}$                  & 42$^{+58}_{-17}$ \\ \hline
~$\chi^2_{\nu}$ &~0.65 ($\frac{43.3}{67}$)~&~0.66 ($\frac{43.8}{66}$)~& ~1.14 ($\frac{114.5}{101}$)~ \\
\hline
\end{tabular}
\end{center}
Parameters: \\
$\dot M$     -- mass accretion rate in the {\it KerrD\/} model   \\
$R_{\rm in}$ -- inner radius of the disk in the {\it KerrD\/} model in $R_{\rm g}=G M/c^2$ 
                 units \\
$k T_{\rm in}$ -- seed photons temperature for Comptonization in keV \\
$\Gamma_e$   -- index of a power law electron energy distribution \\
$\tau$      -- Thomson depth of the Comptonizing cloud \\
$\Omega/2\pi$ -- solid angle of the reprocessor \\
$\xi = 4 \pi F_{\rm X}/n_e$ -- ionization parameter of the reprocessor \\
$R_{\rm refl}$ -- inner radius of the reprocessing disk

\end{table*}

\subsection{State A spectra}

State A is very similar to state B, although somewhat less luminous (\cite{bel00}).
Presence of the non-thermal tail was not explicitly demonstrated in this state,
but given its similarity to state B, we assume the hard tail is indeed given by 
a non-thermal Comptonization and include it in all models.
Initial fits with the non-thermal {\it compPS\/} model
give bad fits ($\chi^2_\nu \approx 3$--$5$). Again, a complex
shape of soft component helps to obtain good fits, with comparable quality of fits
for the various models (Table~\ref{tab:chi2}). The overall spectra are dominated by
the strong soft component, with the harder tail appearing only above $\approx 20$ keV
(Fig.~\ref{fig:44specs}; Table~\ref{tab1}).

The model which includes the Kerr metric disk component gives  comparable
fit quality to the other models with complex soft component. Moreover, the unscattered photons
contribute significantly to the soft component, broadening it compared to what
would be obtained with pure relativistic disk emission (Fig.~\ref{fig:44specs}).
Removing the unscatered photons from the model 
gives clearly unacceptable fits ($\chi^2_\nu\sim 3$), i.e.\ the Kerr disk component with
pure non-thermal Comptonized component cannot describe the data.
In one case (K-45-03) the best overall fit 
is obtained with the blackbody model
while in another case (K-45-02) the Comptonized disk blackbody and the blackbody models provide 
the same quality of fits to the data.
The temperature of the additional blackbody is rather high, $\sim 2$ keV.
It appears that this component is used to adjust the curvature of the Comptonized 
spectrum rather than represents a real spectral component. Clearly, relativistic
effects from a Kerr metric may be considered as contributing to the complexity of the soft component in the
case of K-44-00 observation, but are not sufficient to describe it completely.

Other aspects of the best fit models are
similar to state B spectra: the reprocessed component is present in the data,
it is highly ionized and further smeared (again, with rather large errors).
 An additional narrow gaussian line at 6.4 keV
was also included in the model, since the fit residuals clearly indicated the presence
of such a line. Its equivalent width is $\sim 30$ eV.

Fits of the {\it KerrD\/} model with the non-thermal  tail give $\dot M$ systematically 
lower than those for state B, while $\Rin$ is the same within the fit uncertainty 
(Table~\ref{tab1}).


\subsection{State C spectra}

State C is generally rather harder than either A and B, and the disk component
is less pronounced (\cite{bel00}). 
The non-thermal {\it compPS\/} model gives adequate fits to the data (with our assumed
systematic errors of 1\%).
The $\chi_\nu^{2}$ values for K-44-00, K-45-03, K-05-00, I-42-00 observations are 
$\chi_\nu^{2} =$  1.14, 0.83, 0.87 and 1.04, respectively (Table~\ref{tab:chi2}, model 1; 
Fig.~\ref{fig:44specs}). 

 The disk component can be separated from the total Comptonization component as computed by
{\it compPS\/} by setting the un-scattered fraction of seed photons to 0, and modeling
the soft photons by a separate model. We perform such a procedure for the
K-44-00 and K-45-03 data sets, describing the soft component by the {\it KerrD\/} model. Although
it does not lead to satisfactory fits, it allows to approximate values of physical disk parameters, $\dot M$
and $\Rin$. These are ${\dot M} \approx 2\times 10^{19}\,{\rm g\,s^{-1}}$
and $\Rin \approx 15\,\Rg$. The mass accretion rate is rather high, close to the upper
values for state B. The large inner radius means that the maximum color temperature
is $\approx 0.8$ keV and most of the corresponding luminosity is not seen in the
X--ray band.

In all data sets the reprocessed component is always significantly present. 
Its amplitude is smaller than 1,
it is highly ionized ($\xi \sim 10^4$, but with large errors). Additional broadening of the spectral features
is also required. If this is interpreted as relativistic, the inner radius of the
reflecting disk is $\sim 25$--$100\,\Rg$ (Table~\ref{tab1}).

\section{Discussion}

Our analysis of {\it RXTE\/} data of the micro-quasar \grs reveals all the spectral features 
typically found in high/soft states of accreting black holes (see Gierli\'{n}ski et 
al.\ \cite{gier99}; \.{Z}ycki et al.\ \cite{zyc01}; Zdziarski et al.\ \cite{zdz01}). 
The hard component
is well fit by a non-thermal Comptonization, the soft component is usually complex,
and the reprocessed component is present, highly ionized and further broadened
and smeared.

Clear complex soft component is seen in the high luminosity states A and B.
A number of descriptions of the component are possible:
in addition to the usual disk thermal emission, there may be contribution from hotter
spots on the disk, heated by strong illumination by compact magnetic flares.
This would produce an additional blackbody emission, such as in our model 3.
Alternatively, the hot upper layers of the heated accretion disk may
additionally comptonize the soft photons from below, producing the second,
low temperature comptonized component (see more detailed discussion in, 
e.g., \.{Z}ycki et al.\ 2001).
However, the  relativistic distortions expected from
such a  highly inclined accretion disk ($i\approx 70^\circ$) 
are significant and they cannot be ignored. 
In particular Doppler blueshift and boosting make the
observed temperature higher than the true color disk temperature. However, even assuming
a maximally rotating black hole, we need to include a contribution
from additional  soft photons, in order to explain the spectrum in both state A and B.
Therefore, quantitative determinations of the black hole  angular momentum are not 
possible without constructing a more detailed geometrical model explaining
the origin of all components. However, relativistic effects alone in  Schwarzschild 
metric ($a=0$) would not be able to account for the high apparent peak energy
of the soft component, 
if the mass of $14\,\MSun$ is adopted. In the interpretation of Zdziarski et al.\ 
(\cite{zdz01})
the high apparent peak energy is due to Compton upscattering by a relatively cool,
optically thick plasma (see fig.~3b in their paper), but clearly the Comptonization 
parameters will be changed
if relativistic effects are taken into account in this description.

In state C we generally see strongly comptonized emission 
with small contribution of soft seed photons.
The parameters of the soft component correspond to larger inner disk radius than
in states A and B, confirming qualitatively previous studies of Belloni et al.\ 
(\cite{bel97}).
We note however that quantitative estimates of e.g.\ the inner disk radius
require proper decomposition of spectra into the disk and Comptonized emission.
Models used in many previous papers, i.e.\ {\it diskbb\/} and a power law are clearly
inadequate for such a task. In  particular a power law is not a proper description
of a Comptonization process, if the seed photons are visible in the considered 
energy band, as is clearly the case in GRS~1915+105.
Our modelling suggests inner radii $\approx 15\,\Rg$.
Since in a Kerr metric the dissipation of gravitational energy is strongly concentrated
towards the center, therefore the total accretion power is implied to be much larger
than that directly observed as an X--ray emission. Quantitative estimates would require
knowledge of the black hole spin, but even for a mild value, $a=0.5$, the fraction 
of energy dissipated within $15\,\Rg$ is $\approx 80\%$. Total accretion power
implied would then not be different in state C compared to states A and B, but most
of that power would be used to e.g.\ cause an ejection of plasma.

\begin{acknowledgements}
This research has made use of data obtained through the High Energy Astrophysics
Science Archive Research Center Online Service, provided by the NASA/Goddard Space
Flight Center. Partial support was provided by Polish KBN through grant 2P03D01718.
\end{acknowledgements}

\end{document}